\documentstyle[12pt,prd,aps,epsfig,preprint]{revtex}

\begin{document}
\title{A perturbative treatment for the bound states of the Hellmann potential }
\author{Sameer M. Ikhdair\thanks{%
sikhdair@neu.edu.tr} and \ Ramazan Sever\thanks{%
sever@metu.edu.tr}}
\address{$^{\ast }$Department of Physics, Near East University, Nicosia, North
Cyprus, Mersin-10, Turkey\\
$^{\dagger }$Department of Physics, Middle East Technical University, 06531
Ankara, Turkey.}
\date{\today
}
\maketitle

\begin{abstract}
A new approximation formalism is applied to study the bound states of the
Hellmann potential, which represents the superposition of the attractive
Coulomb potential $-a/r$ and the Yukawa potential $b\exp (-\delta r)/r$ of
arbitrary strength $b$ and screening parameter $\delta $. Although the
analytic expressions for the energy eigenvalues $E_{n,l\text{ }}$ yield
quite accurate results for a wide range of $n,\ell $ in the limit of very
weak screening, the results become gradually worse as the strength $b$ and
the screening coefficient $\delta $ increase. This is because that the
expansion parameter is not sufficiently small enough to guarantee the
convergence of the expansion series for the energy levels.


PACS\ No: 03.65-W; 03.65.Ge; 03.65 Sq
\end{abstract}

\ \ \

\section{Introduction}

\noindent Adamowski [1] has presented a study of the systematics of the
energy eigenvalues of the two-particles interacting via the superposition of
the Coulomb and Yukawa potential (SCYP):

\begin{equation}
V\left( r\right) =-a/r+b\exp (-\delta r)/r,
\end{equation}
where $a$ and $b$ are the strengths of the Coulomb and the Yukawa
potentials, respectively, and $\delta $ is the screening parameter. It is
assumed that $a$ and $\delta $ are positive whereas $b$ can be positive or
negative. The potential in (1) with $b$ positive was first suggested by
Helmann [2,3] \ many years ago and henceforth this potential will be
referred to as the Hellmann potential irrespectively of the sign of $b$. The
Hellmann potential has been used by various authors to represent the
electron-core [4,5] or the electron-ion [6,7] interaction. Varshni and
Shukla [8] used this model potential for alkali hydride molecules. Das and
Chakravarty [9] \ have proposed that such a potential is suitable for the
study of inner-shell ionization problems.

The bound-state energies of the Hellmann potential for various sets of
values of $b$ and $\delta $ have been studied elaborately by Adamowski in a
variational framework using ten variational parameters. The energy
eigenvalues have been predicted very accurately but the calculations involve
extensive computational time and effort. Moreover, compact analytic
expressions far the energy eigenvalues are not obtainable. On the other
hand, Dutt {\it et al} [10] have also been investigated the bound-state
energies as well as the wave functions of this potential using the large-$N$
expansion technique.

In this paper, we study the bound-state properties by applying a new
methodology [11] based on the decompose of the radial Schr\"{o}dinger
equation into two pieces having an exactly solvable part and an additional
piece leading to either a closed analytical solution or an approximate
treatment depending on the nature of the additional perturbed potential. The
application [12,13] of this novel treatment to the different problems in
both, bound and continuum regions, have been proven the success of the
formalism. We demonstrate here how such interaction potential can be simply
treated within the framework of the present formalism.

One of the prime motivations of the present study is to explore the regions
of validity of this approximation formalism for the superposition of
potentials such as the one in (1) which manifests different structures for
various range of values of $b$ and $\delta $. Our calculations reveal that
the degree of accuracy of the predicted eigenvalues vaires appreciably for
different range of values of $b$ and $\delta $ and for different quantum
states. These observations have relevance in the context of applications of
this novel treatment to realistic problems of atomic physics. The other
motivation is that potential (1) with $a=0$ and $b=-\alpha Z$ can be reduced
into the static screened Coulomb potential (SSCP) of the simple form:

\begin{equation}
V\left( r\right) =-(\alpha Z)\exp (-\delta r)/r,
\end{equation}
where $a=(137.037)^{-1}$ is the fine-structure constant and $Z$ is the
atomic number, is often used to compute bound-state normalizations and
energy levels of neutral atoms [14,15,16,17] which have been studied over
the past years. It is known that SSCP yields reasonable results only for the
innermost states when $Z$ is large. However, for the outermost and middle
atomic states, it gives rather poor results. Although the bound- state
energies for the SSCP with $Z=1$ have been studied [18] in the light of the
shifted large-$N$ method. Recently, we have also investigated that this
novel treatment is useful in predicting bound-state energy levels of light
to heavy neutral atoms [17].

The contents of this paper is as follows. In section \ref{TM} we breifly
outline the method with all necessary formulae to perform the current
calculations. In section \ref{AHP} we apply the approach to the
Schr\"{o}dinger equation with Hellmann potential and present the results
obtained analytically for the bound-state energy values upto third
perturbation energy shift. Section \ref{NR} contains the numerical results.
Finally, in section \ref{C} we give our conclusions.

\section{The Method}

\label{TM}For the consideration of spherically symmetric potentials, the
corresponding Schr\"{o}dinger equation, in the bound state domain, for the
radial wave function reads

\begin{equation}
\frac{\hbar ^{2}}{2m}\frac{\psi _{n}^{\prime \prime }(r)}{\psi _{n}\left(
r\right) }=V(r)-E_{n},
\end{equation}
with

\begin{equation}
V\left( r\right) =\left[ V_{0}(r)+\frac{\hbar ^{2}}{2m}\frac{\ell (\ell +1)}{%
r^{2}}\right] +\Delta V(r),
\end{equation}
where $\Delta V(r)$ is a perturbing potential and $\psi _{n}(r)=\chi
_{n}(r)u_{n}(r)$ is the full radial wave function, in which $\chi _{n}(r)$
is the known normalized eigenfunction of the unperturbed Schr\"{o}dinger
equation whereas $u_{n}(r)$ is a moderating function corresponding to the
perturbing potential. Following the prescription of Refs. [11,12,13], we may
rewrite (3) in the form:

\begin{equation}
\frac{\hbar ^{2}}{2m}\left( \frac{\chi _{n}^{\prime \prime }(r)}{\chi _{n}(r)%
}+\frac{u_{n}^{\prime \prime }(r)}{u_{n}(r)}+2\frac{\chi _{n}^{\prime
}(r)u_{n}^{\prime }(r)}{\chi _{n}(r)u_{n}(r)}\right) =V(r)-E_{n}.
\end{equation}
The logarithmic derivatives of the unperturbed $\chi _{n}(r)$ and perturbed $%
u_{n}(r)$ wave functions are given by

\begin{equation}
W_{n}(r)=-\frac{\hbar }{\sqrt{2m}}\frac{\chi _{n}^{\prime }(r)}{\chi _{n}(r)}%
\text{ \ \ and \ \ }\Delta W_{n}=-\frac{\hbar }{\sqrt{2m}}\frac{%
u_{n}^{\prime }(r)}{u_{n}(r)},
\end{equation}
which leads to

\begin{equation}
\frac{\hbar ^{2}}{2m}\frac{\chi _{n}^{\prime \prime }(r)}{\chi _{n}(r)}%
=W_{n}^{2}(r)-\frac{\hbar }{\sqrt{2m}}W_{n}^{^{\prime }}(r)=\left[ V_{0}(r)+%
\frac{\hbar ^{2}}{2m}\frac{\ell (\ell +1)}{r^{2}}\right] -\varepsilon _{n},
\end{equation}
where $\varepsilon _{n}$ is the eigenvalue for the exactly solvable
potential of interest, and

\begin{equation}
\frac{\hbar ^{2}}{2m}\left( \frac{u_{n}^{\prime \prime }(r)}{u_{n}(r)}+2%
\frac{\chi _{n}^{\prime }(r)u_{n}^{\prime }(r)}{\chi _{n}(r)u_{n}(r)}\right)
=\Delta W_{n}^{2}(r)-\frac{\hbar }{\sqrt{2m}}\Delta W_{n}^{\prime
}(r)+2W_{n}(r)\Delta W_{n}(r)=\Delta V(r)-\Delta \varepsilon _{n},
\end{equation}
in which $\Delta \varepsilon _{n}=E_{n}^{(1)}+E_{n}^{(2)}+E_{n}^{(3)}+\cdots
$ is the correction term to the energy due to $\Delta V(r)$ and $%
E_{n}=\varepsilon _{n}+\Delta \varepsilon _{n}.$ If Eq. (8), which is the
most significant piece of the present formalism, can be solved analytically
as in (7), then the whole problem, in Eq. (3) is
\begin{equation}
\left[ W_{n}(r)+\Delta W_{n}(r)\right] ^{2}-\frac{\hbar }{\sqrt{2m}}\left[
W_{n}(r)+\Delta W_{n}(r)\right] ^{\prime }=V(r)-E_{n},
\end{equation}
which is a well known treatment within the frame of supersymmetric quantum
theory (SSQT) [19]. Thus, if the whole spectrum and corresponding
eigenfunctions of the unperturbed interaction potential are known, then one
can easily calculate the required superpotential $W_{n}(r)$ for any state of
interest leading to direct computation of related corrections to the
unperturbed energy and wave function.

For the perturbation technique, we can split the given potential (3) into
two parts. The main part corresponds to a shape invariant potential, Eq.
(7), for which the superpotential is known analytically and the remaining
part is treated as a perturbation, Eq. (8). Therefore, it is obvious that
Hellmann potential can be treated using this prescription. In this case, the
zeroth-order term corresponds to the Coulomb potential while higher-order
terms consitute the perturbation. However, the perturbation term in its
present form cannot be solved exactly through Eq. (8). Thus, one should
expand the functions related to the perturbation in terms of the
perturbation parameter $\lambda $,

\begin{equation}
\Delta V(r;\lambda )=\sum_{i=1}^{\infty }\lambda _{i}V_{i}(r),\text{ \ \ }%
\Delta W_{n}(r;\lambda )=\sum_{i=1}^{\infty }\lambda _{i}W_{n}^{(i)}(r),%
\text{ \ }E_{n}(\lambda )=\sum_{i=1}^{\infty }\lambda _{i}E_{n}^{(i)},
\end{equation}
where $i$ denotes the perturbation order. Substitution of the above
expansions into Eq. (8) and equating terms with the same power of $\lambda $%
\ on both sides up to $O(\lambda ^{4})$ gives

\begin{equation}
2W_{n}(r)W_{n}^{(1)}(r)-\frac{\hbar }{\sqrt{2m}}\frac{dW_{n}^{(1)}(r)}{dr}%
=V_{1}(r)-E_{n}^{(1)},
\end{equation}

\begin{equation}
W_{n}^{(1)}(r)W_{n}^{(1)}(r)+2W_{n}(r)W_{n}^{(2)}(r)-\frac{\hbar }{\sqrt{2m}}%
\frac{dW_{n}^{(2)}(r)}{dr}=V_{2}(r)-E_{n}^{(2)},
\end{equation}

\begin{equation}
2\left[ W_{n}(r)W_{n}^{(3)}(r)+W_{n}^{(1)}(r)W_{n}^{(2)}(r)\right] -\frac{%
\hbar }{\sqrt{2m}}\frac{dW_{n}^{(3)}(r)}{dr}=V_{3}(r)-E_{n}^{(3)},
\end{equation}
\begin{equation}
2\left[ W_{n}(r)W_{n}^{(4)}(r)+W_{n}^{(1)}(r)W_{n}^{(3)}(r)\right]
+W_{n}^{(2)}(r)W_{n}^{(2)}(r)-\frac{\hbar }{\sqrt{2m}}\frac{dW_{n}^{(4)}(r)}{%
dr}=V_{4}(r)-E_{n}^{(4)}.
\end{equation}
Hence, unlike the other perturbation theories, Eq. (8) and its expansion,
Eqs. (11)-(14), give a flexibility for the easy calculations of the
perturbative corrections to energy and wave functions for the $nth$ state of
interest through an appropriately chosen perturbed superpotential.

\section{Application to the Hellmann Potential}

\label{AHP}Considering the recent interest in various power-law potentials
in the literature, we work through the article within the frame of low
screening parameter. In this case, the Hellmann potential can be expanded in
power series of the screening parameter $\delta $ as [13,17,20]

\begin{equation}
V(r)=-\frac{a}{r}+\frac{b}{r}\exp (-\delta r)=-\frac{a}{r}+\frac{b}{r}%
\sum_{i=0}^{\infty }V_{i}(\delta r)^{i},
\end{equation}
where the perturbation coefficients $V_{i}$ are given by

\begin{equation}
V_{1}=+1,\text{ }V_{2}=-1/2,\text{ }V_{3}=1/6,\text{ }V_{4}=-1/24,\text{ }%
\cdots .
\end{equation}
We now apply this approximation method to the Hellmann potential with the
angular momentum barrier

\begin{equation}
V(r)=-\frac{a}{r}+\frac{b}{r}\exp (-\delta r)+\frac{\ell (\ell +1)\hbar ^{2}%
}{2mr^{2}}=\left[ V_{0}(r)+\frac{\ell (\ell +1)\hbar ^{2}}{2mr^{2}}\right]
+\Delta V(r),
\end{equation}
where the first piece is the shape invariant zeroth-order which is an
exactly solvable piece corresponding to the unperturbed Coulomb potential
with $V_{0}(r)=-(a-b)/r$ while $\Delta V(r)=-b\delta +(b\delta
^{2}/2)r-(b\delta ^{3}/6)r^{2}+(b\delta ^{4}/24)r^{3}-\cdots $ is the
perturbation term. The literature is rich with examples of particular
solutions for such power-law potentials employed in different fields of
physics, for recent applications see Refs. [21,22]. At this stage one may
wonder why the series expansion is truncated at a lower order. This can be
understood as follows. It is widely appreciated that convergence is not an
important or even desirable property for series approximations in physical
problems. Specifically, a slowly convergent approximation which requires
many terms to achieve reasonable accuracy is much less valuable than the
divergent series which gives accurate answers in a few terms. This is
clearly the case for the Hellmann problem [23]. However, it is worthwhile to
note that the main contributions come from the first three terms. Thereby,
the present calculations are performed up to the third-order involving only
these additional potential terms, which suprisingly provide highly accurate
results for small screening parameter $\delta .$

\subsection{Ground State Calculations $\left( n=0\right) $}

In the light of Eq. (7), the zeroth-order calculations leading to exact
solutions can be carried out readily by setting the ground-state
superpotential and the unperturbed exact energy as

\begin{equation}
W_{n=0}\left( r\right) =-\frac{\hbar }{\sqrt{2m}}\ \frac{\ell +1}{r}+\sqrt{%
\frac{m}{2}}\frac{(a-b)}{(\ell +1)\hbar },\text{ \ \ }E_{n}^{(0)}=-\frac{%
m(a-b)^{2}}{2\hbar ^{2}(n+\ell +1)^{2}},\text{ \ \ \ }n=0,1,2,....
\end{equation}
and from the literature, the corresponding normalized Coulomb bound-state
wave function [13,17,24]

\begin{equation}
\chi _{n}^{(C)}(r)=N_{n,l}^{(C)}r^{\ell +1}\exp \left[ -\beta r\right]
\times L_{n}^{2\ell +1}\left[ 2\beta r\right] ,
\end{equation}
in which $N_{n,l}^{(C)}=\left[ \frac{2m(a-b)}{\left( n+\ell +1\right) \hbar
^{2}}\right] ^{\ell +1}\frac{1}{(n+\ell +1)}\sqrt{\frac{m(a-b)n!}{\hbar
^{2}(n+2\ell +1)!}}$ is a normalized constant,\ \ $\beta =\frac{m(a-b)}{%
\left( n+\ell +1\right) \hbar ^{2}}$ and $L_{n}^{k}\left( x\right)
=\sum_{m=0}^{n}(-1)^{m}\frac{(n+k)!}{\left( n-m\right) !(m+k)!m!}x^{m}$ is
an associate Laguarre polynomial function [25].

For the calculation of corrections to the zeroth-order energy and
wavefunction, one needs to consider the expressions leading to the first to
the third-order perturbation given by Eqs. (11)--(14). Multiplication of
each term in these equations by $\chi _{n}^{2}(r)$, and bearing in mind the
superpotentials given in Eq. (6), one can obtain the straightforward
expressions for the first-order correction to the energy and its
superpotential:
\begin{equation}
E_{n}^{(1)}=\int_{-\infty }^{\infty }\chi _{n}^{2}(r)\left( \frac{b\delta
^{2}}{2}r\right) dr,\text{ }W_{n}^{(1)}\left( r\right) =\frac{\sqrt{2m}}{%
\hbar }\frac{1}{^{X_{n}^{2}(r)}}\int^{r}\chi _{n}^{2}(x)\left[ E_{n}^{(1)}-%
\frac{b\delta ^{2}}{2}x\right] dx,\
\end{equation}
and for the second-order correction and its superpotential:

\[
E_{n}^{(2)}=\int_{-\infty }^{\infty }\chi _{n}^{2}(r)\left[ -\frac{b\delta
^{3}}{6}r^{2}-W_{n}^{(1)}\left( r\right) W_{n}^{(1)}\left( r\right) \right]
dr,\text{ }
\]
\begin{equation}
W_{n}^{(2)}\left( r\right) =\frac{\sqrt{2m}}{\hbar }\frac{1}{^{X_{n}^{2}(r)}}%
\int^{r}\chi _{n}^{2}(x)\left[ E_{n}^{(2)}+W_{n}^{(1)}\left( r\right)
W_{n}^{(1)}(x)+\frac{b\delta ^{3}}{6}x^{2}\right] dx,
\end{equation}
and for the third-order correction and its superpotential:

\[
E_{n}^{(3)}=\int_{-\infty }^{\infty }\chi _{n}^{2}(r)\left[ \frac{b\delta
^{4}}{24}r^{3}-W_{n}^{(1)}\left( r\right) W_{n}^{(2)}\left( r\right) \right]
dr,\text{ }
\]
\begin{equation}
W_{n}^{(3)}\left( r\right) =\frac{\sqrt{2m}}{\hbar }\frac{1}{^{X_{n}^{2}(r)}}%
\int^{r}\chi _{n}^{2}(x)\left[ E_{n}^{(3)}+W_{n}^{(1)}(x)W_{n}^{(2)}(x)-%
\frac{b\delta ^{4}}{24}x^{3}\right] dx,
\end{equation}
for any state of interest. The above expressions calculate $W_{n}^{(1)}(r),$
$W_{n}^{(2)}(r)$\ and $W_{n}^{(3)}(r)$ explicitly from the energy
corrections $E_{n}^{(1)},$ $E_{n}^{(2)}$ and $E_{n}^{(3)}$ respectively,
which are in turn used to calculate the moderating wave function $u_{n}(r).$

Thus, using Eqs. (20)-(22), one finds the zeeroth order energy shift and
their moderating superpotentials, for $a\neq b,$ as

\[
E_{0}^{(1)}\ =\frac{\hbar ^{2}b(3N_{0}^{2}-L)}{4(a-b)m}\delta ^{2},\text{ \ }%
W_{0}^{(1)}(r)=\frac{\hbar bN_{0}\delta ^{2}}{2\sqrt{2m}(a-b)}r,
\]

\[
E_{0}^{(2)}=-\frac{\hbar ^{4}bN_{0}^{2}\left( 5N_{0}^{2}-3L+1\right) }{%
12(a-b)^{2}m^{2}}\delta ^{3}-\frac{\hbar ^{6}b^{2}N_{0}^{4}\left(
5N_{0}^{2}-3L+1\right) }{16(a-b)^{4}m^{3}}\delta ^{4},
\]

\[
W_{0}^{(2)}(r)=-\frac{\hbar N_{0}\left[ (a-b)mr+\hbar ^{2}N_{0}N_{1}\right] %
\left[ 3\hbar ^{2}b^{2}N_{0}^{2}\delta +4mb(a-b)^{2}\right] \delta ^{3}}{24%
\sqrt{2m}m^{2}(a-b)^{4}}r,
\]

\[
E_{0}^{(3)}\ =\frac{\hbar ^{6}bN_{0}^{2}\left( 5N_{0}^{2}-3L\right) \left(
5N_{0}^{2}-3L+1\right) }{96(a-b)^{3}m^{3}}\delta ^{4}+\frac{\hbar
^{8}b^{2}N_{0}^{4}\left( 5N_{0}^{2}-3L+1\right) \left( 9N_{0}^{2}-5L\right)
}{48(a-b)^{5}m^{4}}\delta ^{5}
\]
\begin{equation}
+\frac{\hbar ^{10}b^{3}N_{0}^{6}\left( 5N_{0}^{2}-3L+1\right) \left(
9N_{0}^{2}-5L\right) }{64(a-b)^{7}m^{5}}\delta ^{6},
\end{equation}
where $N_{0}=\left( \ell +1\right) ,$ $N_{1}=\left( \ell +2\right) $ and $%
L=\ell (\ell +1).$ Therefore, the analytical expressions for the lowest
energy and full radial wave function of the Hellmann potential are then
given by

\[
E_{n=0,\ell }=E_{n=0}^{(0)}-b\delta
+E_{0}^{(1)}+E_{0}^{(2)}+E_{0}^{(3)}+\cdots ,\text{ }\psi _{n=0,\ell
}(r)\approx \chi _{n=0,\ell }^{(C)}(r)u_{n=0,\ell }(r),
\]

\begin{equation}
u_{n=0,\ell }(r)\approx \exp \left( -\frac{\sqrt{2m}}{\hbar }\int^{r}\left(
W_{0}^{(1)}\left( x\right) +W_{0}^{(2)}\left( x\right) \right) dx\right) .
\end{equation}
Hence, the explicit form of the full wave function in (24) for the ground
state is

\begin{equation}
\psi _{n=0,\ell }(r)=\left[ \frac{2m(a-b)}{(\ell +1)\hbar ^{2}}\right]
^{\ell +1}\frac{1}{(\ell +1)}\sqrt{\frac{(a-b)m}{\hbar ^{2}(2\ell +1)!}}%
r^{\ell +1}\exp (P(r)),
\end{equation}
with $P(r)=\sum_{i=2}^{3}p_{i}r^{i}$ is a polynomial of third order having
the following coefficients:
\begin{equation}
p_{2}=\frac{bN_{0}\delta ^{2}}{4(a-b)}\left[ \frac{N_{1}\hbar ^{2}c}{m}-1%
\right] ,\text{ \ \ }p_{3}=\frac{1}{6}bc\delta ^{2},
\end{equation}
in which

\begin{equation}
c=\frac{N_{0}\delta }{12m(a-b)^{3}}\left[ 3\hbar ^{2}bN_{0}^{2}\delta
+4m(a-b)^{2}\right] .
\end{equation}

\subsection{Excited state calculations $(n\geq 1)$}

The procedures applied in the calculations of the ground states becomes
extremely cumbersome in the description of radial excitations when nodes of
wavefunctions are taken into account, in particular during the higher order
calculations. Although several attempts have been made to bypass this
difficulty and improve calculations in dealing with excited states, (cf.
e.g. [26], and the references therein) within the frame of supersymmetric
quantum mechanics.

Using Eqs. (6), (18) and (19), the superpotential $W_{n}(r)$ which is
related to the excited states can be readily calculated by means of Eqs.
(20)-(22). Therefore, the energy shifts in the first excited state, with $%
a\neq b,$ are:

\[
E_{1}^{(1)}=\frac{\hbar ^{2}b(3N_{1}^{2}-L)}{4m(a-b)}\delta ^{2},\text{\ \ }%
W_{1}^{(1)}(r)=\frac{\hbar bN_{1}\delta ^{2}}{2\sqrt{2m}(a-b)}r,
\]

\[
\ E_{1}^{(2)}=-\frac{\hbar ^{4}bN_{1}^{2}\left( 5N_{1}^{2}-3L+1\right) }{%
12(a-b)^{2}m^{2}}\delta ^{3}-\frac{\hbar ^{6}b^{2}N_{1}^{4}\left(
5N_{1}^{2}-3L+1\right) }{16(a-b)^{4}m^{3}}\delta ^{4},
\]

\[
W_{1}^{(2)}(r)=-\frac{\hbar N_{1}\left[ (a-b)mr+\hbar ^{2}N_{1}N_{2}\right] %
\left[ 3\hbar ^{2}b^{2}N_{1}^{2}\delta +4mb(a-b)^{2}\right] \delta ^{3}}{24%
\sqrt{2m}m^{2}(a-b)^{4}}r,
\]

\[
E_{1}^{(3)}\ =\frac{\hbar ^{6}bN_{1}^{2}\left( 5N_{1}^{2}-3L\right) \left(
5N_{1}^{2}-3L+1\right) }{96(a-b)^{3}m^{3}}\delta ^{4}+\frac{\hbar
^{8}b^{2}N_{1}^{4}\left( 5N_{1}^{2}-3L+1\right) \left( 9N_{1}^{2}-5L\right)
}{48(a-b)^{5}m^{4}}\delta ^{5}
\]
\begin{equation}
+\frac{\hbar ^{10}N_{1}^{6}\left( 5N_{1}^{2}-3L+1\right) \left(
9N_{1}^{2}-5L\right) }{64(a-b)^{7}m^{5}}\delta ^{6},
\end{equation}
Therefore, the approximated energy value of the Hellmann potential
corresponding to the first excited state is\

\begin{equation}
E_{n=1,\ell }=E_{1}^{(0)}-b\delta
+E_{1}^{(1)}+E_{1}^{(2)}+E_{1}^{(3)}+\cdots .
\end{equation}
The related radial wavefunction can be expressed in an analytical form in
the light of Eqs (20), (21) and (24), if required. The appromation used in
this work would not affect considerably the sensitivity of the calculations.
On the other hand, it is found analytically that our investigations put
forward an interesting hierarchy between $W_{n}^{(1)}(r)$ terms of different
quantum states in the first order with $a\neq b$ after circumventing the
nodal difficulties elegantly,\ \ \

\begin{equation}
W_{n}^{(1)}(r)=\frac{\hbar bN_{n}\delta ^{2}}{2\sqrt{2m}(a-b)}r,
\end{equation}
where $N_{n}=(n+\ell +1).$ Therefore, for the second excited state $\left(
n=2\right) $ leads to the first-order energy shift and superpotential

\begin{equation}
\ E_{2}^{(1)}=\frac{\hbar ^{2}b(3N_{2}^{2}-L)}{4m(a-b)}\delta ^{2},\text{\ }%
W_{2}^{(1)}(r)=\frac{\hbar bN_{2}\delta ^{2}}{2\sqrt{2m}(a-b)}r,
\end{equation}
where $N_{2}=\left( \ell +3\right) .$ Thus, the use of $W_{2}^{(1)}(r)$ in
Eq.(21) gives the energy shift and supersymmetric potential in the
second-order with $a\neq b$ as\

\[
\ E_{2}^{(2)}=-\frac{\hbar ^{4}bN_{2}^{2}\left( 5N_{2}^{2}-3L+1\right) }{%
12(a-b)^{2}m^{2}}\delta ^{3}-\frac{\hbar ^{6}b^{2}N_{2}^{4}\left(
5N_{2}^{2}-3L+1\right) }{16(a-b)^{4}m^{3}}\delta ^{4},
\]
\bigskip

\begin{equation}
W_{2}^{(2)}(r)=-\frac{\hbar N_{2}\left[ (a-b)mr+\hbar ^{2}N_{2}N_{3}\right] %
\left[ 3\hbar ^{2}b^{2}N_{2}^{2}\delta +4mb(a-b)^{2}\right] \delta ^{3}}{24%
\sqrt{2m}(a-b)^{4}m^{2}}r.
\end{equation}
Finally, we obtain the third-order energy shift as

\[
E_{2}^{(3)}\ =\frac{\hbar ^{6}bN_{2}^{2}\left( 5N_{2}^{2}-3L\right) \left(
5N_{2}^{2}-3L+1\right) }{96(a-b)^{3}m^{3}}\delta ^{4}+\frac{\hbar
^{8}b^{2}N_{2}^{4}\left( 5N_{2}^{2}-3L+1\right) \left( 9N_{2}^{2}-5L\right)
}{48(a-b)^{5}m^{4}}\delta ^{5}
\]
\begin{equation}
+\frac{\hbar ^{10}b^{3}N_{2}^{6}\left( 5N_{2}^{2}-3L+1\right) \left(
9N_{2}^{2}-5L\right) }{64(a-b)^{7}m^{5}}\delta ^{6}.
\end{equation}
Therefore, the approximated energy eigenvalue of the Hellmann potential
corresponding to the second excited state is\

\begin{equation}
E_{n=2,\ell }=E_{2}^{(0)}-b\delta
+E_{2}^{(1)}+E_{2}^{(2)}+E_{2}^{(3)}+\cdots .
\end{equation}
In general, from the supersymmetry, we find out the $nth$ state energy
shifts together with their supersymmetric potentials with $a\neq b$ as

\[
E_{n}^{(1)}\ =\frac{\hbar ^{2}b\left[ 3N_{n}^{2}-L\right] }{4(a-b)m}\delta
^{2},\text{\ }W_{n}^{(1)}(r)=\frac{\hbar bN_{n}\delta ^{2}}{2\sqrt{2m}(a-b)}%
r.
\]

\[
E_{n}^{(2)}=-\frac{\hbar ^{4}bN_{n}^{2}\left[ 5N_{n}^{2}-3L+1\right] }{%
12(a-b)^{2}m^{2}}\delta ^{3}-\frac{\hbar ^{6}b^{2}N_{n}^{4}\left[
5N_{n}^{2}-3L+1\right] }{16(a-b)^{4}m^{3}}\delta ^{4},
\]

\[
W_{n}^{(2)}(r)=-\frac{\hbar N_{n}\left[ (a-b)mr+\hbar ^{2}N_{n}N_{n+1}\right]
\left[ 3\hbar ^{2}b^{2}N_{n}^{2}\delta +4mb(a-b)^{2}\right] \delta ^{3}}{24%
\sqrt{2m}m^{2}(a-b)^{4}}r,
\]

\begin{eqnarray}
E_{n}^{(3)}\ &=&\frac{\hbar ^{6}bN_{n}^{2}\left[ 5N_{n}^{2}-3L\right] \left[
5N_{n}^{2}-3L+1\right] }{96(a-b)^{3}m^{3}}\delta ^{4}+\frac{\hbar
^{8}b^{2}N_{n}^{4}\left[ 5N_{n}^{2}-3L+1\right] \left[ 9N_{n}^{2}-5L\right]
}{48(a-b)^{5}m^{4}}\delta ^{5}  \nonumber \\
&&+\frac{\hbar ^{10}b^{3}N_{n}^{6}\left( 5N_{n}^{2}-3L+1\right) \left(
9N_{n}^{2}-5L\right) }{64(a-b)^{7}m^{5}}\delta ^{6},
\end{eqnarray}
where $N_{n+1}=(n+l+2).$ Consequently, the total energy for the nth state is

\begin{equation}
E_{n,\ell }=E_{n}^{(0)}-b\delta +E_{n}^{(1)}+E_{n}^{(2)}+E_{n}^{(3)}+\cdots .
\end{equation}

\section{Numerical Results}

\label{NR}For the numerical work, we take $a=2$ and thus our $b$ is to be
identified as the corresponding rescaled parameter in Adamowski's paper.
Consequently, our energy eigenvalues are like those obtained by Adamowski.
In Tables 1 and 2, we compute the binding energies, $-E_{n,l},$ of the
lowest-lying states (from $1s$ up to $4f$) for various values of $b=\pm
1,-2,-4,-10$ as functions of the screening parameter $\delta $ obtained from
the analytic expressions given in Eqs. (18), (35) and (36). The dependance
of the energy levels on $b$ is shown in Table 3 for the states $1s$ up to $%
3d.$ The results for the higher excited states (from $5s$ up to $7i$) are
presented in Table 4. The energy eigenvalues $1s-6h$ for the attractive
Yukawa potential with $\delta =0.1$ as functions of $b$ are shown in Table
5. The predicted results are then compared with the accurate energy
eigenvalues [1] obtained by Adamowski using a high precision variational
technique.

Therefore, we display our results in Tables 1 and 2 only for some sets of
values of $b$ and $\delta .$ Although we do not present here all the energy
eigenvalues considered by Adamowski, our calculation reveals certain
interesting features of this approximation method.

The present calculations show that the binding spectra of the Hellmann
potential possess the following properties.

(i) For low and strong coupling of $b$ in Yukawa part, the energy
eigenvalues obtained from the perturbation method are in good agreement with
the variational results for low values of the screening parameter $\delta $.
Obviously, when $\delta $ is small the Coulomb field character prevails and
the perturbation method has been adjusted, of course, to that. However, the
results become gradually worse as the screening parameter $\delta $ becomes
large. Appreciable discrepancy of our results from the variational
calculations occurs almost for most states if $\delta >0.2.$ We suspect that
this happen because the perturbative parts of potential becomes so shallow
and its minimum shifts appreciably from the minimum of the true potential.
For certain values of $\delta $, the perturbation potentials for some states
becomes so shallow that the expansion for the energy seies becomes divergent
in the sense that higher perturbation terms in Eq.(35) dominate over the
unperturbed term in Eq.(18) and cousequently one gets anomalous results.

(ii) For almost most strongly attractive (negative) $b$ but small $\delta $,
it is possible to determine the binding energy eigenvalues for $1s$ up to $%
4f $ states.

(iii) For a given $n,$ the results of the energy eigenvalues, $E_{n\ell }$
increase with increasing $\ell $ if the Yukawa potential is attractive ($b<0$%
), and $E_{n\ell }$ decrease with increasing $\ell $ if the Yukawa potential
is repulsive ($b>0$); i.e., for $\ell >\ell ^{^{\prime }},$ $E_{n\ell
}-E_{n\ell ^{^{\prime }}}>0$ or $E_{n\ell }-E_{n\ell ^{^{\prime }}}<0$ if $%
b<0$ or $b>0,$ respectively. This is found to be consistent with the level
ordering theorem of Grosse and Martin [27]. The $s$ levels are mostly split
off from the hydrogenlike levels $E_{n}^{H}$ [Eq.(18)] (downwards for $b<0$
and upwards for $b>0).$ The energy eigenvalues $E_{n\ell }$ approach $%
E_{n}^{H}$ if $n$ and $\ell $ increase. The shift of the energy levels with
respect to $E_{n}^{H}$ is due to an influence of the finite range Yukawa
potential in Eq.(2).

(iv) For the attractive Yukawa potential there exist some values of $b$ and $%
\delta $ for which the energy levels with larger $n$ and smaller $\ell $
become lower than the levels with smaller $n$ and larger $\ell ,$ i.e., $%
E_{n\ell }\leq E_{n^{\prime }\ell ^{^{\prime }}}$ if $n>n^{^{\prime }}\geq 3$
and $\ell <\ell ^{^{\prime }}.$

(v) For the repulsive Yukawa potential there are some values of the strength
$b$ and the screening parameter $\delta $ for which the energy eigenvalues
for larger $n$ and $\ell $ become lower than those for smaller $n$ and $\ell
,$ i.e., $E_{n\ell }\leq E_{n^{\prime }\ell ^{^{\prime }}}$ if $%
n>n^{^{\prime }}\geq 2$ and $\ell >\ell ^{^{\prime }}.$

On the other hand, in Tables 6 and 7, we present the ground and excited
energies (in units such that $\hbar ^{2}/2m=1$) for several states $\ell
=0,1,2$ calculated for the potential in Eq.(2) and compare them with other
works [28]. One finds that our results are remarkably good for the $\ell =0$
ground state. As $\ell $ increases for a given $b,$ the error increases.
However as $b$ increases, the relative error decreases rapidly. This is
particularly useful when $b$ is large. For large $b,$ the calculated
energies compare very well with high precision numerical calculation
presented by other works [11,12,28].

\section{Conclusions}

\label{C} The detailed analysis of the results in terms of various domains
of parameters $b$ and $\delta $ of the Hellmann potential reveals a few
important facts concerning the application of the perturbed formalism. In
the present study the discrete energy eigenvalues for the Hellmann potential
have been calculated as functions of the strength $b$ and the screening
parameter $\delta $ of the Yukawa potential. For $b=0$ the spectrum is given
by Eq. (18). The energy eigenvalues $E_{n\ell }$ for the Hellmann potential
are shifted upwards or downwards with respect to the hydrogenlike levels,
Eq. (18), if $b>0$ or $b<0,$ respectively. The absolute values of deviations
of $E_{n\ell }$ from $E_{n}^{H}$ decrease with increasing quantum number $%
\ell ,$ which results in the corresponding ordering of the energy levels for
a given $n.$ This is due to the influence of the finite range Yukawa
potential (YP), which decreases with increasing $\ell .$

The properties of the energy spectrum for the Hellmann potential obtained in
the present work have many analogies in atomic, solid-state, and quark
physics. Property (i), giving the order of the energy levels $E_{n\ell }$
with the same $n,$ dependent on $\ell $ and on the sign of the YP, has an
application to such systems as an exciton and a bound polaron in polar
semiconductors and ionic crystals. These systems consist of two oppositely
charged particles interacting with themselves through a polarizable medium.
The energy levels of both the systems exhibit this case, although the total
effective potential is even more complicated than the Hellmann potential,
being a linear combination of the Coulomb potential and an additional
potential, which is a sum of two Yukawa potentials with different strengths
and screening parameters, and an exponential potential. However, the net
contribution of the additional potential is negative for the exciton and
positive for the bound polaron and one of the Yukawa potentials dominates at
small distances [29,30,31].

Another system, having the energy levels ordered similarly to those for the
Hellmann potential with $b>0,$ is the quarkonium (the bound state of heavy
quark-antiquark pair). For each $n$ the energy levels of the quarkonium
system increase with increasing $\ell .$ This can be explained in the frames
of the simple model, assuming that the quark interact via the potential
being the superposition of the attractive Coulomb potential and the positive
linear potential (Cornell potential) [32].

\acknowledgments Sameer M. Ikhdair wishes to thank the president of
the Near East University Dr. Suat G\"{u}nsel, vice presidents Prof.
Dr. \c{S}enol Bekta\c{s} and Prof. Dr. Fakhreddin Mamedov for the
partial financial support. He also dedicates this work to his family
members for their love and assistance.This research was partially
supported by the Scientific and Technical Research Council of
Turkey.

\bigskip

\begin{table}[tbp]
\caption{Energy eigenvalues, $-E_{n\ell },$ of states $1s-4f$ for Hellmann
potential as a function of screening parameter $\protect\delta .$ Energy
eigenvalues are given in units of Table 1.}
\begin{tabular}{lllllll}
State%
\mbox{$\backslash$}%
$\delta $ & $0.001$ & $0.005$ & $0.01$ & $0.05$ & $0.1$ & $0.2$ \\
\tableline$b=-10$ &  &  &  &  &  &  \\
$1s$ & $35.99$ & $35.95$ & $35.9001$ & $35.5031$ & $35.0124$ & $34.0489$ \\
$2s$ & $8.99$ & $8.95012$ & $8.90050$ & $8.51527$ & $8.0482$ & $7.18656$ \\
$2p$ & $8.99$ & $8.95010$ & $8.90042$ & $8.51025$ & $8.04037$ & $7.15688$ \\
$3s$ & $3.99001$ & $3.95028$ & $3.90112$ & $3.52703$ & $3.10435$ & $2.39157$
\\
$3p$ & $3.99001$ & $3.95026$ & $3.90103$ & $3.52508$ & $3.09701$ & $2.36532$
\\
$3d$ & $3.99001$ & $3.95022$ & $3.90087$ & $3.52120$ & $3.08240$ & $2.31330$
\\
$4s$ & $2.24002$ & $2.2005$ & $2.15197$ & $1.79676$ & $1.42667$ & $0.88467$
\\
$4p$ & $2.24002$ & $2.20048$ & $2.15189$ & $1.79490$ & $1.41993$ & $0.86351$
\\
$4d$ & $2.24002$ & $2.20043$ & $2.15173$ & $1.79119$ & $1.40652$ & $0.82123$
\\
$4f$ & $2.24001$ & $2.20037$ & $2.15148$ & $1.78565$ & $1.38656$ & $0.757901$%
\end{tabular}
\end{table}

\bigskip \bigskip
\begin{table}[tbp]
\caption{Energy eigenvalues,$-E_{n\ell },$ of states $1s-3d$ as function of
the strength $b$ of the Hellmann potential for $\protect\delta =0.01.$
Energy eigenvalues are given in units of Table 1.}
\begin{tabular}{lllllll}
$b$%
\mbox{$\backslash$}%
State & $1s$ & $2s$ & $2p$ & $3s$ & $3p$ & $3d$ \\
\tableline$1$ & $0.259852$ & $0.071928$ & $0.0720203$ & $0.036557$ & $%
0.0366436$ & $0.0368131$ \\
$0.5$ & $0.567450$ & $0.145431$ & $0.145463$ & $0.067079$ & $0.0671090$ & $%
0.0671683$ \\
$0.2$ & $0.811983$ & $0.204435$ & $0.204446$ & $0.091858$ & $0.0918682$ & $%
0.0918884$ \\
$0$ & $1$ & $0.25$ & $0.25$ & $0.111$ & $0.111$ & $0.111$ \\
$-0.2$ & $1.20801$ & $0.300553$ & $0.300545$ & $0.132562$ & $0.132553$ & $%
0.132537$ \\
$-0.5$ & $1.55753$ & $0.385743$ & $0.385723$ & $0.168871$ & $0.168852$ & $%
0.168815$ \\
$-1$ & $2.24005$ & $0.552697$ & $0.552664$ & $0.240435$ & $0.240404$ & $%
0.240341$ \\
$-2$ & $3.98007$ & $0.980297$ & $0.980248$ & $0.425103$ & $0.425055$ & $%
0.424959$ \\
$-5$ & $12.2$ & $3.01293$ & $3.01286$ & $1.31206$ & $1.31199$ & $1.31185$ \\
$-10$ & $35.9$ & $8.9005$ & $8.90042$ & $3.90112$ & $3.90103$ & $3.90087$ \\
$-20$ & $120.8$ & $30.051$ & $30.0505$ & $13.2460$ & $13.2456$ & $13.2454$%
\end{tabular}
\end{table}

\bigskip

\begin{table}[tbp]
\caption{Energy eigenvalues, $-E_{n\ell },$ of states $5s-7i$ as function of
the strength $b$ of the Hellmann potential for $\protect\delta =0.01.$
Energy eigenvalues are given in units of Table 1.}
\begin{tabular}{lllllll}
State%
\mbox{$\backslash$}%
$b$ & $+1$ & $-1$ & $-2$ & $-4$ & $-8$ & $-10$ \\
\tableline$5s$ & $0.0165535$ & $0.081148$ & $0.141758$ & $0.322393$ & $%
0.922921$ & $1.34306$ \\
$5p$ & $0.0166884$ & $0.0811191$ & $0.141713$ & $0.322331$ & $0.922844$ & $%
1.34298$ \\
$5d$ & $0.0169435$ & $0.0810614$ & $0.141624$ & $0.322208$ & $0.922692$ & $%
1.34282$ \\
$5f$ & $0.017289$ & $0.0809751$ & $0.141490$ & $0.322023$ & $0.922463$ & $%
1.34258$ \\
$5g$ & $0.0176805$ & $0.0808607$ & $0.141312$ & $0.321776$ & $0.922158$ & $%
1.34225$ \\
$6s$ & $0.0100105$ & $0.0540995$ & $0.093579$ & $0.213386$ & $0.618604$ & $%
0.904359$ \\
$6p$ & $0.0103092$ & $0.054072$ & $0.0935360$ & $0.213326$ & $0.618530$ & $%
0.904281$ \\
$6d$ & $0.0108763$ & $0.054017$ & $0.0934503$ & $0.213206$ & $0.618380$ & $%
0.904123$ \\
$6f$ & $0.011651$ & $0.0539351$ & $0.0933218$ & $0.213027$ & $0.618155$ & $%
0.903887$ \\
$6g$ & $0.0125417$ & $0.0538266$ & $0.0931512$ & $0.212787$ & $0.617856$ & $%
0.903572$ \\
$6h$ & $0.013427$ & $0.0536924$ & $0.0929390$ & $0.212489$ & $0.617482$ & $%
0.903178$ \\
$7s$ & $0.0028019$ & $0.0380167$ & $0.0648977$ & $0.148193$ & $0.435796$ & $%
0.640565$ \\
$7p$ & $0.0020524$ & $0.0379905$ & $0.0648565$ & $0.148135$ & $0.435723$ & $%
0.640488$ \\
$7d$ & $0.0006123$ & $0.0379383$ & $0.0647744$ & $0.148018$ & $0.435576$ & $%
0.640333$ \\
$7f$ & $0.0014009$ & $0.0378606$ & $0.0646516$ & $0.147845$ & $0.435356$ & $%
0.640101$ \\
$7g$ & $0.0038107$ & $0.0377579$ & $0.0644887$ & $0.147613$ & $0.435064$ & $%
0.639792$ \\
$7h$ & $0.0063819$ & $0.0376312$ & $0.0642864$ & $0.147325$ & $0.434698$ & $%
0.639405$ \\
$7i$ & $0.0088204$ & $0.0374816$ & $0.0640455$ & $0.146980$ & $0.434260$ & $%
0.638942$%
\end{tabular}
\end{table}

\bigskip

\begin{table}[tbp]
\caption{Energy eigenvalues, $-E_{n\ell },$ of states $1s-6h$ for $\protect%
\delta =0.1.$ Energy eigenvalues are given in units of Table 1.}
\begin{tabular}{lllllll}
State%
\mbox{$\backslash$}%
$b$ & $-5$ & $-8$ & $-10$ & $-20$ & $-30$ & $-50$ \\
\tableline$1s$ & $11.7605$ & $24.2118$ & $35.0124$ & $119.014$ & $253.014$ &
$671.014$ \\
$2s$ & $2.60282$ & $5.49595$ & $8.0482$ & $28.3034$ & $61.0555$ & $164.057$
\\
$2p$ & $2.59637$ & $5.48852$ & $8.04037$ & $28.2947$ & $61.0463$ & $164.048$
\\
$3s$ & $0.946371$ & $2.0766$ & $3.10435$ & $11.562$ & $25.5672$ & $70.2385$
\\
$3p$ & $0.94056$ & $2.06971$ & $3.09701$ & $11.5536$ & $25.5583$ & $70.2292$
\\
$3d$ & $0.929052$ & $2.05602$ & $3.0824$ & $11.5367$ & $25.5406$ & $70.2106$
\\
$4s$ & $0.405183$ & $0.92834$ & $1.42667$ & $5.76553$ & $13.2138$ & $37.4734$
\\
$4p$ & $0.400245$ & $0.922129$ & $1.41993$ & $5.75749$ & $13.2052$ & $%
37.4643 $ \\
$4d$ & $0.390444$ & $0.909778$ & $1.40652$ & 5.74145 & 13.188 & 37.4461 \\
$4f$ & $0.37593$ & $0.89143$ & $1.38656$ & 5.71747 & 13.1624 & 37.4189 \\
$5s$ & $0.185308$ & $0.44099$ & $0.700006$ & 3.14653 & 7.56577 & 22.3833 \\
$5p$ & $0.181729$ & $0.43572$ & $0.694047$ & 3.13896 & 7.55755 & 22.3745 \\
$5d$ & $0.174538$ & $0.425204$ & $0.682168$ & 3.12385 & 7.54112 & 22.3568 \\
$5f$ & $0.16367$ & $0.40949$ & $0.664442$ & 3.10128 & 7.51654 & 22.3304 \\
$5g$ & $0.149027$ & $0.38865$ & $0.640982$ & 3.07135 & 7.48388 & 22.2952 \\
$6s$ & $0.099768$ & $0.213846$ & $0.348659$ & 1.78538 & 4.56702 & 14.2628 \\
$6p$ & $0.097267$ & $0.209807$ & $0.343761$ & 1.77836 & 4.5592 & 14.2543 \\
$6d$ & $0.092169$ & $0.201681$ & $0.333949$ & 1.76435 & 4.54358 & 14.2372 \\
$6f$ & $0.084289$ & $0.189376$ & $0.319193$ & 1.74341 & 4.5202 & 14.2116 \\
$6g$ & $0.073342$ & $0.172749$ & $0.299444$ & 1.71562 & 4.48915 & 14.17775
\\
$6h$ & $0.058954$ & $0.151614$ & $0.274642$ & 1.68109 & 4.45051 & 14.1351
\end{tabular}
\end{table}

\bigskip

\begin{table}[tbp]
\caption{Some bound state energies, $-E_{n\ell },$ for neutral atoms. }
\begin{tabular}{lllll}
$b=-Ze^{2}$ & State & Present & Calculated [28] & Exact [28] \\
\tableline-4 & 1s & 3.25647 & 3.219965 & 3.250536 \\
& 2p & 0.37172 &  &  \\
& 3d & 0.003891 &  &  \\
-8 & 1s & 14.4581 & 14.419973 & 14.457119 \\
& 2s & 2.61756 &  &  \\
& 2p & 2.58365 & 2.433176 & 2.583677 \\
& 3d & 0.54034 &  &  \\
-16 & 1s & 60.8590 & 60.819285 & 60.859039 \\
& 2s & 13.0276 & 13.027315 & 13.03259 \\
& 2p & 12.9910 & 12.837459 & 12.991055 \\
& 3s & 4.39692 & 4.372037 & 4.405697 \\
& 3p & 4.36357 & 4.348041 & 4.388576 \\
& 3d & 4.29724 &  &  \\
-24 & 1s & 139.2590 & 139.22008 & 139.259362 \\
& 2s & 31.4314 & 31.431281 & 31.43595 \\
& 2p & 31.3938 & 31.238464 & 31.393815 \\
& 3s & 11.7014 & 11.699808 & 11.70926 \\
& 3p & 11.6662 & 11.665284 & 11.683877 \\
& 3d & 11.5959 & 11.245640 & 11.595949 \\
& 4s & 5.05037 & 5.044185 & 5.0590 \\
& 4p & 5.01801 & 5.013466 & 5.05410 \\
& 4d & 4.95352 & 4.951600 & 5.00855 \\
& 5s & 2.21653 & 2.203309 & 2.22372 \\
& 5p & 2.18793 & 2.177040 & 2.241432 \\
& 5d & 2.13080 & 2.124124 & 2.24278
\end{tabular}
\end{table}

\bigskip

\begin{table}[tbp]
\caption{Some bound state energies, $-E_{n\ell },$ for neutral atoms. }
\begin{tabular}{llllllll}
$b=-Ze^{2}$ & $1s$ & $2s$ & $2p$ & $3s$ & $3p$ & $4s$ & $4p$ \\
$-13$ & $39.7088$ & $8.18765$ & $8.15172$ & $2.57025$ & $2.53816$ & $0.82329$
& $0.79663$ \\
$-36$ & $316.86$ & $74.0341$ & $73.9958$ & $29.3129$ & $29.2763$ & $13.9316$
& $13.897$ \\
$-79$ & $1544.51$ & $374.5$ & $374.461$ & $158.088$ & $158.05$ & $82.6361$ &
$82.599$%
\end{tabular}
\end{table}

\bigskip

\end{document}